\documentstyle[12pt]{article}

\newcommand{\beq}{\begin{equation}}
\newcommand{\eeq}{\end{equation}}

\begin{document}
\def\lag{\langle}
\def\rag{\rangle}

\title{Properties of Interfaces in the\\
two and three dimensional\\
Ising Model\\[1in]}

\author{B.A. Berg$^{1,2}$, U. Hansmann$^{1,2}$ and T. Neuhaus$^3$
}
\vfill
\footnotetext[1]{{\em
 Department of Physics, The Florida State University, Tallahassee,
 FL 32306, USA}}
\footnotetext[2]{{\em
 Supercomputer Computations Research Institute Tallahassee, The Florida
 State University, Tallahassee, FL 32306, USA}}
\footnotetext[3]{{\em
 Fakult\"at f\"ur Physik, Universit\"at Bielefeld,
 D-W-4800 Bielefeld, FRG}}

\maketitle
\begin{abstract}

To investigate order-order interfaces, we perform multimagnetical Monte
Carlo simulations of the $2D$ and $3D$ Ising model.
Following Binder we extract the interfacial free energy from the
infinite volume limit of the magnetic probability density. Stringent
tests of the numerical methods are performed by reproducing with high
precision exact $2D$ results. In the physically more interesting
$3D$ case we estimate the amplitude $F^s_0$ of the critical interfacial
tension $F^s = F^s_0 t^\mu$ to be $F^s_0 = 1.52 \pm 0.05$. This result is
in good agreement with a previous MC calculation by Mon, as well as with
experimental results for related amplitude ratios. In addition, we
study in some details the shape of the magnetic probability density
for temperatures below the Curie point.
\end{abstract}

\newpage

\section{Introduction}

Over the last fifty years, and presumably longer, there has been
continuous interest in the properties of interfaces in $2D$ and $3D$
Ising models. In the $2D$ case a number of remarkable exact results
could be derived \cite{Onsager,RotW,Shlo}, whereas
in $3D$ Monte Carlo (MC) simulations \cite{Bin1,Bin2,Mon,Mey}
have become the dominant method. Of
particular interest are amplitude ratios \cite{Fisk,Stauffer}
which involve the amplitude $F^s_0$ of the critical $3D$ interfacial
tension, because they can be compared with a variety of experimental
results \cite{Mod,Gie} for fluids which are supposed to populate the
Ising model universality class. Additional interest in $F^s_0$ is created
by the fact that it controls the non-universality of the critical
wetting transition, a subject in which one has seen continuous interest
over the last years \cite{Fish1}.

Let us introduce the notation. Spins $s_i = \pm 1$ are defined on  sites
of a square lattice of volume $V = L^D$ with periodic boundary conditions
and the symbol $<i,j>$ is used to denote nearest neighbours. The partition
function of the $D$-dimensional Ising model is given by
\begin{equation}
Z\ =\ \sum_{\rm configurations} \exp ( - \beta H) .
\end{equation}
The Ising Hamiltonian $H$
\begin{equation}
H\ =\ H_I\ -\ h M ~~~{\rm with}~~~ H_I = - \sum_{<i,j>} s_i s_j
{}~~{\rm and}~~ M = \sum_i s_i
\label{ising}
\end{equation}
contains the nearest neighbour interaction term $H_I$ and a term
which couples the external magnetic field $h$ to the magnetization.
We denote the magnetic probability density by $P_L (M)$. Often it is
preferable to use the magnetization density $m=M/L^D$ and
\begin{equation}
p_L (m)\ =\ P_L (M) .
\end{equation}
When convenient arguments are suppressed altogether.
Let us consider the case of a vanishing magnetic field $h=0$.
In the broken region $\beta > \beta_c$ the magnetic probability
densities $P_L (M)$ are double peaked. We denote the positions of
the maxima by $\pm M^{\max}_L$. As the model is globally $Z(2)$
symmetric, we exploit the symmetry $P_L (M) = P_L (-M)$,
define $P_L^{\max} = P_L(M_L^{\max}) = P_L (-M_L^{\max})$,
and confine our discussion to $M>0$. For $M_L<M_L^{\max}$ the
probability density takes then its minimum at
$P_L^{\min} = P_L (0)$.

The surface tension $F^s$ is defined as the the free energy per unit
area of the interface between coexisting phases. Here we consider
order-order interfaces in the broken region $\beta > \beta_c$. The
asymptotic behaviour for $\beta\to\beta_c$ is
\begin{equation}
F^s\ =\ F^s_0 t^\mu + F^s_1 t^{\mu +1} ,
\label{tens0}
\end{equation}
where $t=(1-\beta_c/\beta)$ is the reduced temperature. Widom's scaling
relation \cite{Widom}
\begin{equation}
\mu\ =\ (D-1) \nu ,
\label{widom}
\end{equation}
connects the exponent $\mu$ with the critical exponent $\nu$ of the
correlation length. In $2D$ $\mu=\nu=1$, together with $F^s_0 = 1.762...$
and $F^s_1 = 1.216...$ follows from exact results \cite{Onsager}. In
$3D$ approximate methods have to be used, and $\nu = 0.630$ is in agreement
with \cite{Zin,Alv1} as well as consistent with many more estimates
of the extensive literature. Therefore, we use $\mu = 1.26$ in $3D$
throughout this paper. When combined with other amplitudes, $F^s_0$
enters into a number of universal ratios \cite{Fisk,Stauffer}.
The other amplitudes are fairly accurately known, see for instance
\cite{Liu}, and the uncertainty in the ratios stems mainly from $F^s_0$.

Numerical calculations of interface tensions have remained subtle and a
large number of methods have been tried. Old estimates of $F^s_0$
\cite{Bin2} led to amplitude ratios which are in disagreement with the
experimental results for fluids \cite{Mod,Gie},
whereas Mon \cite{Mon} obtained consistency by calculating the excess
energy between a system with an interface imposed by an antiperiodic
boundary condition and a system without an interface with periodic
boundary conditions. Another promising approach is to calculate the
interfacial tension from correlation functions \cite{Jan,Mey}.
In this paper we turn back to the approach of Binder \cite{Bin1,Bin2}
and combine it with a new simulation method \cite{our1,our2}.

The interface tension $F^s$ can be defined from the finite size scaling
(FSS) behaviour of $P_L^{\min} / P_L^{\max}$, which for large $L$ takes
the form
\begin{equation}
{P^{\min}_L \over P^{\max}_L}\ =\
{\rm const}\ L^p\ \exp \left[ -2 L^{D-1} F^s (1+a_1L^{-1}+a_2L^{-2}
+ O(L^{-3})) \right]
\label{tens1}
\end{equation}
on lattices with periodic boundary conditions. Physically this
definition assumes that at values of the mean magnetization $M = 0$ a
rectangular domain, enclosed by two interfaces and spanning the lattice
via the periodic boundary condition, is formed. Interface shapes other
than planar, see for instance Figure 7 in section~6, are due to
capillary waves which have to be included in the definition of $F^s$.
It is convenient to introduce the function
\begin{equation}
\Psi_L (M)\ =\ L^{-(D-1)} ~\ln {P_L^{\max} \over P_L (M) }
{}~~~{\rm and}~~~ \psi_L (m)\ =\ \Psi_L (M) .
\label{tens2}
\end{equation}
The finite lattice interfacial tension is then defined as
\begin{equation}
F^s_L\ =\ {1\over 2} \Psi_L (0)\ =\ {1\over 2} \psi_L (0) ,
\label{tens3}
\end{equation}
and $F^s = \lim_{L\to\infty} F^s_L$. It follows that the infinite
volume limit
\begin{equation}
\psi (m)\ =\ \lim_{L\to\infty} \psi_L (m)
\label{tens4}
\end{equation}
exists and takes values in the range $[0,F^s]$.

Originally, the calculation of $F^s$ via (\ref{tens1}) suffered from
a problem of principle. To get a representative sample of interfaces
one has to generate many configurations with $M=0$ on large lattices.
But, in a canonical simulation this is precisely impossible because
of (\ref{tens1}). This equation means that the relevant configurations
are exponentially suppressed, implying a slowing down of the simulation
$\sim \exp (+2L^{D-1} F^s)$. Recently this difficulty was overcome by
the proposal to perform the MC simulation for a multimagnetical ensemble
\cite{our2}, a natural extension of the multicanonical ensemble
introduced in \cite{our1}. The slowing down becomes then reduced to a
power law close to $\sim V^2$.
Details of this approach are given in the next section.

In section 3 the numerical methods are introduced which we employ to
cope with equation (\ref{tens0}) on the basis of our multimagnetical
MC data. Section 4 summarizes our $2D$ and section 5 our $3D$ results.
Besides the surface tension, this includes tunneling times and
data for the magnetic susceptibility. The $2D$ Ising model serves
essentially as a testing ground for the numerical
approach, whereas the $3D$ results are of more physical interest.
Section 6 discusses the shape of the magnetical probability
density in greater details. A brief summary is given in our final
section 7.
\hfill\break

\section{The multimagnetical ensemble}

The standard MC algorithm would only sample configurations corresponding
to $P^{\min}_L$ if one could generate of the order $1/P^{\min}_L$
or more statistically independent configurations. We overcome this
difficulty by sampling configurations with a multimagnetical weight factor
\begin{equation}
{\cal P}^{mm}_L (M)\ \sim\ n(M)^{-1}
{}~~~{\rm for}~~~ 0 \le M \le M^{\max}_L.
\label{muma0}
\end{equation}
Here $n(M)$ is the magnetical density of of states (fixed temperature
$\beta^{-1}$). For any reasonable approximation the resulting
multimagnetic probability density will be approximately flat.
For reasons of physical intuition the parameterization
\begin{equation}
{\cal P}^{mm}_L (M)\ = \exp ( \alpha_L (M) + h_L (M) \beta M - \beta H_I )
{}~~~{\rm for}~~~ 0 \le M \le V
\label{muma1}
\end{equation}
is useful. For constant $\alpha,\ h$ the weight factor
$\exp (\alpha + h\beta M - \beta H)$ describes a canonical ensemble in an
external magnetic field $h$.
The function $h_L (M)$ defines an $M$-dependent effective external magnetic
field such that the resulting multimagnetical probability density is flat
for $M\le M_L^{\max}$. As $M$ changes in steps of two, the definition of
$h_L (M)$ for $2\le M\le M_L^{\max}-2$ is
\begin{equation}
\exp \left[ 2 \beta h_L(M) \right]\ =\ {P_L (M+2) \over P_L (M)} .
\label{muma2}
\end{equation}
Including now $M=0$ and $M\ge M_L^{\max}$ leads to
\begin{equation}
h_L (M)\ =\ \cases{0 ~~{\rm for}~ M=0 ~{\rm and}~ M\ge M^{\max} ;  \cr
2^{-1} \beta^{-1} L^{D-1} \left[ \Psi_L (M) - \Psi_L (M+2) \right]
                                     ~{\rm elsewhere}~ (M>0). \cr}
\label{muma3}
\end{equation}
For $M < 0$, $h_L(M)$ is defined through $h_L(M)=-h_L(-M)$. The purpose of
the function $\alpha_L (M)$ is to ensure continuity of the multimagnetical
weight factor. This leads to the recursion relation
\begin{equation}
\alpha_L (M+2)\ =\ \alpha_L (M) + \beta \left[ h_L(M)-h_L(M+2) \right]\ (M+2),
\ \alpha_L (0) = 0 .
\label{muma4}
\end{equation}
Once $h_L (M)$ is given, $\alpha_L (M)$ follows automatically.
The standard Markov process is well-suited to generate
configurations which are in equilibrium with respect to this
multimagnetical distribution. The $h=0$ canonical probability density
$P_L (M)$ is obtained from $P_L^{mm} (M)$ by:
\begin{equation}
P_L (M)\ =\ c\ P_L^{mm} (M)\ \exp ( - \alpha_L (M) - h_L (M) \beta M ).
\label{muma5}
\end{equation}
The constant $c$ is obtained by imposing the appropriate
normalization on $P_L (M)$, which we choose $P_L^{\max} = 1$ for
the purposes of this paper.

How do we get now the multimagnetical function $h(M)$? For small systems
$P_L (M)$ can accurately be calculated by performing standard heatbath
or Metropolis simulations, and $h_L (M)$ follows directly from
(\ref{tens2},\ref{muma3}). On larger systems we get $h_L (M)$ by making
every time a FSS prediction of $\Psi_L (M)$ from the already controlled
smaller systems. To
optimize the parameters we may perform a second run and in some cases further
runs to increase our statistics. We now explain details of the FSS
prediction. The outlined method is simple and worked sufficiently well
for our practical purposes, but the approach is by no means unique and
can likely be improved.

{}From the definition of the interface tension we know that $\psi_L (0)$
scales like
\begin{equation}
\psi_L (0)\ =\ 2  F^s + {a^0 \over L^{D-1}} ,
\label{muma6}
\end{equation}
where we neglect $\ln (L)$ terms and $a^0$ is an unknown constant.
Further, $m^{\max}_L = M^{\max}_L / L^D$ is a function of the lattice
size and scales like
\begin{equation}
m^{\max}_L =  m^{\max} \left( 1 + {c \over L^{D -1}} \right) .
\label{muma7}
\end{equation}
We employ this equation for re-adjusting the magnetization $m_L$ in general,
and use a scaling law of the form (\ref{muma6}) for the entire function
$\psi_L (m)$:
\begin{equation}
\psi_L (m_L) = \psi (m) + {a^m \over L^{D-1} } ,
\label{muma8}
\end{equation}
where $\psi (m)$ and $a^m$ are two unknown, $m$-dependent constants.
To calculate the multimagnetical parameters for a new lattice size we just
take our two largest lattices, start from the $m_L$ on the smaller
lattice, determine the readjusted magnetizations on the larger lattices
and compute for this set of points $\psi_L (m)$ for the new lattice.
A linear interpolation between these points gives us our final
approximation of $\psi_L (m)$. Using $\Psi (M) = \psi (m)$, the new
multimagnetic function $h_L (m)$ follows again from (\ref{muma3}).
It may be noted that due to the choice of our normalization $a^m = 0$
at $m_L^{\max}$ for all $L$. Since we only need an approximate estimate
of $h_L (M)$, it seems to be harmless that $\Psi_L (M)$ is a noisy estimate.
It is our experience that the noise averages out over sufficient distances
in $m$. In Figure~1 we give an example of an un-reweighted
distribution which we get by this method.

As before \cite{our1,our2} we define the tunneling time $\tau_L$ as
the average number of updates needed to get from a configuration
with $M=M^{\min}=-M^{\max}$ to a configuration with $M = M_L^{\max}$
and back.
Due to (\ref{muma0}) we expect that the standard Markov process leads
to a one dimensional random walk like behaviour of the magnetization
$M$. Because of $M_L^{\max} \sim V$ the slowing down becomes then
reduced to a $V^2$ power law. Our data in the subsequent sections confirm
this up to a small correction, which indicates that the random walk
behaviour is not entirely perfect.
The deeper we are in the broken region, the more more dramatic is the
improvement which we expect. Because $F^s$ increases with $\beta$,
this follows from (\ref{tens1}).
\hfill\break

\section{Numerical analysis of the surface tension}

To calculate $F^s_L$ via (\ref{tens1}-\ref{tens3}) requires the
maxima and the minimum of the probability density $P_L (M)$.
To determine the minimum causes no problem because we know
$P^{\min}_L = P_L (0)$. To estimate $P_L^{\max}$ we exploit the exactly
known symmetry $P_L (M) = P_L (-M)$: $P^{\max}_L$ is estimated by looking
for the absolute maximum of the distribution on one side, taking the
reflected value on the other side, and averaging over both possibilities.

The final estimate of $F^s$ requires a FSS extrapolation towards
$L=\infty$. Equation (\ref{tens1}) converts into the fit
\begin{equation}
F^s_L\ =\ F^s + {a \over L^{D-1}} + {b\ \ln (L) \over L^{D-1}}
              + {c \over L^{D}} .
\label{fit1}
\end{equation}
Here $c/L^D$ takes care of the $L^{-2}$ corrections to $F^s$ in (\ref{tens1}).
However, the accuracy of our $F^s_L$ data will not allow a four parameter
fit. The obvious three parameter fit (III$_1$) is obtained by setting
$c=0$. Further, it is numerically difficult to include the $a/L^{D-1}$
and the $b \ln (L)/L$ term. Therefore, the three parameter fits defined
by $b=0$ (III$_2$) and $a=0$ (III$_3$) are also considered. Corresponding
two parameter fits (II$_2$) and (II$_3$) follow by demanding $c=0$ in
addition. Although interactions (correlations) between the two interfaces
fall off exponentially with their distance, and hence the system size,
for small systems they may be more important then the $L^{-2}$ correction
to $F^s$ in (\ref{tens1}). Consequently one may try the fit
\begin{equation}
F^s_L\ =\ F^s + {a \over L^{D-1}} + {b\ \ln (L) \over L^{D-1}}
              + c\ e^{-d L}
\label{fit2}
\end{equation}
with either $a=0$ or $b=0$.

One likes to include data from as many lattices as possible,while keeping
the fit self-consistent. The systematic approach towards this fitting problem
begins with fitting the data from all the lattices and monitoring
$\chi^2$ of this fit. The fit is accepted if the obtained $\chi^2$
value results into a reasonable likelihood (here we choose $\ge$ 10~\% )
that the discrepancies between the fit and the data points are due to
statistical fluctuations. If this is not the case, the smallest lattice
is supposed to suffer from strong finite size corrections of a form
not included in the fit. Omitting the smallest lattice, the fit is
carried out again. The whole procedure is repeated until an acceptable,
called self-consistent, fit is obtained (or until one runs out of
sufficiently many data points).

The statistical error of $F^s_L$ has a somewhat peculiar finite size
behaviour. According to our discussion of the previous section we
assume that the multicanonical simulation slows down like $L^{(2+\eta)D}$
with $\eta>0$ small. In this paper we give the statistics in sweeps. One
sweeps updates each spin on the lattice once.
For a constant number of MC sweeps we have
\begin{equation}
{ \triangle P_L^{\min} \over P_L^{\min} }\ \sim\ L^{(1+\eta)D/2},
\label{eb1}
\end{equation}
because the number of independent entries of $P_L (0)$ decreases
$\sim L^{-(1+\eta)D}$. Now (remember our normalization $P_L^{\max} =1$)
\begin{equation}
- \ln \left[ P_L^{\min} \left( 1 \pm {\triangle P_L^{\min} \over P_L^{\min} }
    \right) \right]\ \approx\ 2L^{(D-1)}F^s \pm
    {\triangle P_L^{\min} \over P_L^{\min} }
\label{eb2}
\end{equation}
and, consequently,
\begin{equation}
\left( \triangle F^s_L \right)^2\ \sim\ L^{(\eta -1) D+2}\
\approx\ L^{2-D} .
\label{eb3}
\end{equation}
This equation implies that in
$D=2$ a constant number of sweeps keeps the $F^s_L$ error bar
approximately constant, and in $D=3$ this may even be achieved
with a decreasing statistics ($\sim 1/\sqrt{L}$ for $\eta=0$).
Of course, the statistics has to be chosen such that on the
largest lattice still a sufficiently large number of tunneling events
({\it i.e} independent data) is obtained. Otherwise a meaningful
statistical analysis is impossible. Twenty or more tunneling events
is a a good number for practical purposes and numbers as small as
four do still allow some kind of reasonable analysis \cite{Alv2}.

We pursue two ways to estimate the amplitude $F^s_0$ defined by
(\ref{tens0}). For a sufficiently small reduced temperature
\begin{equation}
F^s_{01}\ :=\ F^s t^{-\mu}
\label{ampl1}
\end{equation}
approximates $F^s_0$ well. On the other hand, when we keep the data
time series re-weighting techniques \cite{Fer,Alv2} allow to calculate
$F^s (\beta)$ in a neighbourhood of the simulation point. This allows
another estimate of $F^s_0$:
\begin{equation}
F^s_{02}\ :=\ \lim_{L\to\infty} {\hat F}^s_L ~~~{\rm with}~~~
{\hat F}^s_L = {d F^s_L \over d t^\mu} = \lim_{\epsilon\to 0}
{F^s_L (t+\epsilon) - F^s_L (t) \over (t+\epsilon)^\mu - t^\mu} .
\label{ampl2}
\end{equation}
Estimates (\ref{ampl1}) and (\ref{ampl2}) differ by the next term in
the expansion (\ref{tens0}):
\begin{equation}
F^s_1\ =\ \mu\ t^{-1} (F^s_{02}-F^s_{01})
\label{ampl3}
\end{equation}
In principle one may use this to estimate $F^s_1$ and, subsequently,
use (\ref{tens0}) to improve the estimate (\ref{ampl1}) of $F^s_0$.
In practice the estimate $F^s_{02}$ provides only a consistency check,
because $F^s_{02}$ turns out to be too noisy to compete with $F^s_{01}$.
\hfill\break

\section{$2D$ estimates}

For the two dimensional Ising model we performed multimagnetical
simulations at the critical temperature
$\beta = {\beta}_c = \ln (1+\sqrt{2})/2 = 0.44068...$, at $\beta = 0.47$
and 0.5. The results are collected in Tables 1--3.
In each run additional $200,000$ initial sweeps without measurements
were performed for reaching equilibrium with respect to the multimagnetical
distribution. To give an example, Figure 2 depicts the $\beta =0.5$
magnetic probability densities $p_L (m)$ for the range $L=10-100$.
Beyond $F^s_L$, $\hat F^s_L$ and $\tau_L$, as introduced in the previous
sections, another quantity included in Tables 1--3 is the magnetic
susceptibility
\begin{equation}
\chi_L\ =\ V \sum_m m^2 p_L (m) .
\label{sus1}
\end{equation}
In the large $L$ limit its leading term will approach the exactly
known \cite{Yang} mean magnetization squared: $(m^0)^2$. Recent
progress in FSS theory raises interest \cite{Bor} in the subleading
terms, which even in $2D$ are not known analytically.

For all $\beta$ values self-consistent three parameter fits
(\ref{fit1}-III$_1$) are possible. The obtained parameters $b$ are
compatible with zero, and the appropriate two parameter fits
(\ref{fit1}-II$_2$) do not decrease the lattice range over
which self-consistency is achieved: $L=14-100$ ($\beta = \beta_c$),
$L=20-100$ ($\beta = 0.47$), and $L=30-1002$ ($\beta = 0.5$).
Including now the $L^{-2}$ term, {\it i.e.} turning to fit
(\ref{fit1}-III$_2$), increases the self-consistency range
considerably: $L=4-100$ ($\beta = \beta_c$), $L=10-100$
($\beta = 0.47$), and $L= 8-100$ ($\beta = 0.5$). Together with
the $F^s_L$ data these fits are depicted in Figure 3, and we rely
on them for our final $F^s$ estimates. Exact results follow
from Onsager's \cite{Onsager} equation
\begin{equation}
F^s\ =\  2\beta - \ln \left[ {1+e^{-2\beta} \over 1-e^{-2\beta}} \right],
{}~~~(\beta \ge \beta_c) .
\label{on1}
\end{equation}
In Tables 1--3 we have included our $L=\infty$ estimates as well
as the exact values. Good agreement is found in all cases.

To calculate the amplitude $F^s_0$ we employ equations (\ref{ampl1})
and (\ref{ampl2}). We list in Tables 1--3 the corresponding numerical
estimates $F^s_{01}$ and $F^s_{02}$ together with their exact values
which follow from Onsager's formula (\ref{on1}). It is seen that
$F^s_{02}$ is systematically larger than $F^s_{01}$. According to
(\ref{ampl3}) this implies $F^s_1 > 1$. Equation (\ref{tens0})
defines then a third, improved approximation $F^s_{03}$ of $F^s_0$
which is even smaller than $F^s_{01}$. This may be illustrated with
exact results: $F^s_{03}=1.76 $ for $\beta =0.47$ and $F^s_{03}=1.75$
for $\beta=0.5$ are obtained. Both values are good approximations of the
exact results $F^s_0=1.74$. The slight improvement for $\beta =0.5$ is
accidentally caused by next order terms.
Unfortunately, the numerical estimates of
$F^s_{02}$ from the $F^s_L$ turn out to be too noisy to make this
a feasible practical method. Still, the increase of the $F^s_{01}$
estimates with $t$ as well as the $F^s_{02}$ give correctly the trend.
In fact the direct two parameter fit (\ref{tens0}) of our two rather
accurate $F^s$ estimates give $F^s_0 = 1.765 \pm 0.003$.
In conclusion, the employed methods pass well testing versus the exactly
known $2D$ Ising model results.

To estimate the slowing down of the multimagnetical simulations
we fit the $\tau_L$ values of Tables 1--3 to the form
$\tau_L \sim L^{2+2\eta}$, corresponding to the suspected
slowing down $\sim V^{2+\eta}$ in updates. We find
$\eta = 0.012 $ ($\beta = \beta_c$),
$\eta = 0.045 $ ($\beta = 0.47$), and
$\eta = 0.048 $ ($\beta = 0.5$). The actual improvement is,
of course, most impressive for $\beta = 0.5$, as the
exponential slowing down of the canonical simulation is there
strongest. See \cite{our2} for a detailed comparison with standard
heat bath simulations at this $\beta$ value. At $\beta = \beta_c$
the standard simulation does not slow down anymore exponentially.
Still we find improvement by a constant factor $\approx 3$, and
this situation is depicted in Figure 4.
\hfill\break

\section{$3D$ estimates}

We performed simulations at $\beta = 0.227$, 0.232 and 0.2439. The
corresponding temperatures are all well above the roughening temperature
$\beta_r \approx 1.85\ \beta_c \approx 0.41$ \cite{Mon2}.
Our results are collected in Tables 4--6. As in the 2D case 200,000
additional, initial sweeps were performed in each run for reaching
equilibrium. Again, we give an example for the magnetic probability
densities. Figure 5 illustrates $\beta = 0.2439$ for $L=8-32$.

In Figure 6 we display the effective tensions as functions of the
lattice size together with asymptotic fits. We notice that finite size
effects play a more important role in three than two dimensions and are
more complicated, too. The non-monotone behaviour shows that it is
necessary to use large enough lattices to estimate the interface tension.
For each $\beta$ value we could only generate few lattices of sufficiently
large size. Consequently, we find that our $F^s_L$ data are only well
suited for two parameter fits. In contrast to $2D$, it turns out that the
$\ln (L)/L$ term seems to be relevant. The two parameter fit
(\ref{fit1}-II$_3$) is preferred by the data when compared with
(\ref{fit1}-II$_2$), presumably both terms would finally
contribute. The self-consistency ranges are $L=24-32$ ($\beta = 0.227$),
$L=20-32$ ($\beta =0.232$), and $L=24-32$ ($\beta = 0.2439$). We include
the thus obtained estimates in Tables 4--6. The accuracy of our data
does not allow to determine the exponent $p$ in (\ref{tens1}) reliably.
Obviously, the situation is
far less satisfactory than in $2D$. Larger lattices are desirable, but
could not be simulated with our present computational resources.
To get a handle on the order of magnitude of the expected systematic
errors, we tried to extend the self-consistency ranges by
including further fit parameters (\ref{fit1}), (\ref{fit2}).
With one exception these fits are either unstable or have a bad
goodness of fit. The exception is the $\beta = 0.232$ point. The
fit (\ref{fit2}) with $a=0$ leads to the self-consistency range
$L=6-32$ and $F^s = 0.03044 \pm 0.00025$. By comparison with the
results reported in Table 5, one may argue that the results of
Tables 4--6 are presumably afflicted by systematic errors of
2\% to 4\%, thus exceeding the statistical errors.

The value $\beta = 0.232$ was chosen, because it enables a comparison.
In \cite{Mey} cluster improved estimators
were used to calculate correlations for $L=8-14$ in a cylindrical geometry.
Fitting the obtained tunneling mass gaps yields $F^s = 0.03034 \pm 0.00015$
for the surface tension. This is in remarkably good agreement with our
fit (\ref{fit2}) of the range $L=6-32$, although this might be accidental.
In any case, one seems to be on the save side by allowing for the discussed
systematic error. All our $F^s$ values are much higher that the old estimates
of \cite{Bin2}, which had to rely on far too small lattices.

To estimate the amplitude $F^s_0$ we first calculate $F^s_{01}$
(\ref{ampl1}) and $F^s_{02}$ (\ref{ampl2}),
as in the previous section, and include the results in
Tables 4--6. In contrast to the $2D$ case, no clear trend of $F^s_{01}$
with increasing $t$ is observed. In view of this, we tend to attribute
their discrepancies to be due to the systematic errors in taking the
asymptotic limits $\lim_{L\to\infty} F^s_L (\beta )$. Consequently,
a fair estimate is obtained by averaging over our three values and
allowing for about 3\% systematic errors. This gives
\begin{equation}
F^s_0\ =\ 1.52 \pm 0.05 .
\label{fs0}
\end{equation}
Surprisingly, our $3D$ results for $F^s_{02}$ suffer less from statistical
noise than in $2D$. As before the $F^s_{02}$ are systematically higher
than the corresponding $F^s_{01}$. This indicates that a further
downward correction of (\ref{fs0}) might be necessary. However,
presently involved ambiguities are too severe to dare such a
correction. Instead one should first try to push the multimagnetical
$3D$ Ising model calculations towards larger lattices, an enterprise which
will become feasible with the next generation(s) of fast workstations.

Again, we perform power law fits of the $\tau_L$ data to estimate the
slowing down of the multimagnetical simulations. We fit the $\tau_L$
of Tables 4--6 to the form $\tau_L \sim L^{3+3\eta}$, corresponding to
the suspected slowing down $\sim V^{2+\eta}$ in updates. We find
$\eta = 0.001$ ($\beta = 0.227$),
$\eta = 0.006$ ($\beta = 0.232$), and
$\eta = 0.030$ ($\beta = 0.2439$).
The real improvement is most dramatic for $\beta = 0.2439$. For the
largest lattice lattice ($L=32$) it is about
$\exp ( 2\times 0.074\times 32^2) \approx 10^{66}$.
\hfill\break

\section{The shape of the probability distributions}

The behavior of the magnetic probability density at small values of
the magnetization is related to the existence and properties of
droplets and domains of one of the pure phases floating in a medium
of the opposite phase. On a given lattice properties of domains will
strongly depend on the chosen boundary conditions, which in our case
are periodic. To proceed we shortly review $2D$ results of \cite{RotW,Shlo}.
The analytical form of $\psi (m)$, defined by (\ref{tens4}), can be given
in terms of three temperature dependent parameters $m^0$, $w$ and $m^c$:
\begin{equation}
\psi(M) ~ =  ~ \cases{
w ~ \sqrt{m^0-m^c} = {\rm const} = 2F^s ~~{\rm for}~~ -m^c\le m\le m^c; \cr
{}~ \cr
w ~ \sqrt{m^0 - \mid m \mid} ~~{\rm for}~~ \mid m \mid \ge m^c . \cr}
\label{psia}
\end{equation}
There exists a finite interval in the absolute value of
the magnetization, where the function $\psi$ is constant, namely
twice the value of the interface tension along one of the main
axes of the lattice. At the value of $m=m^c$ the function $\psi$ develops
a cusp. Following a square root behavior, $\psi$ then approaches
zero at the value of the mean magnetization $m^0$.

Let us outline the qualitative physical picture behind the above formula.
At magnetization $m \approx 0$ half of the spins belong to a domain
with a characteristic magnetization corresponding to one of the pure phases,
e.g. to the state characterized by $+m^0$. Consequently the other half of
spins are contained in a domain structure with typical magnetization $-m^0$.
In between these domains a surface is formed, and dominant contributions to
the partition function at $m \approx 0$ arise from configurations with a
minimal excess free surface energy. These are configurations where a domain
strip is closed via the periodic boundary conditions of the lattice. In
Figure 7 we display a typical configuration contributing at $m \approx 0$ on
the $100^2$ lattice. One clearly observes a strip with sizable fluctuations
added. Such configurations allow translation of one interface without
changing the energy or the number of states. As the magnetization is
changed under such a translation one finds a constant behavior of $\psi$
in a finite interval $\mid m \mid <m^c$. At $m=m^c$ yet another mechanism
takes place. It becomes more favorable for the system to form a droplet
- or bubble - of one pure phase floating in a sea of the opposite phase.
This happens when the surface energy of a droplet is smaller than the
surface energy of a domain. The system enters a region of configuration
space where single or multiple droplets dominate. Figures of such
configurations are given in \cite{our2}.

These statements can be made quantitative for the $2D$ Ising model. Let
us consider the point $\beta =0.5$.
Onsager's \cite{Onsager} interface tension
determines $w ~ \sqrt{m^0-m^c} = 2F^s = 0.45612$ for $\beta =0.5$, and
due to Yang \cite{Yang} $m^0=0.9113$ for $\beta=0.5$.
The remaining free parameter is the
cusp value of the magnetization $m^c$. Its value follows from the Wulf
construction \cite{Wul} of classical equilibrium shapes of droplets.
Relying on analytic results of Rottmann and Wortis \cite{RotW} we display
in Figure 8
equilibrium shapes of droplets for several temperatures below $T_c$.
At very low temperature a quadratic shape is assumed  with small rounding
effects at its corners (at zero temperature the shape is exactly quadratic).
Closer to the transition point the droplet
shapes take a circular form, while exactly at the transition point
due to rotational invariance the droplet shape is a circle.
At $\beta=0.5$ the droplet shape exhibits
only tiny deviations from a circular shape.
When the surface energy of a classical droplet equals the surface energy
of the rectangular domain $2L F^s$, the transition from the droplet
phase to a percolated state takes place.
For a circular droplet shape one finds the relation
\begin{equation}
m^c~=~m^0 ~ \cases{ 1 - 2/\pi  ~~~{\rm for}~~  D=2, \cr
1 - 4/ (3 \sqrt{\pi}) ~~~{\rm for}~~ D=3 ,\cr}
\label{circle}
\end{equation}
while for a quadratic shape $m^c = m^0/2 $, and $m^c = m^0/3$ for a
cubic shape.

In Figure 2 we already presented our numerical $\beta =0.5$
probability distributions $p_L(m)$ on lattices ranging from $10^2$ to
$100^2$. As predicted by theory we observe with increasing lattice size
the unfolding of a flat region at values of the magnetization close
to zero. In order to compare the theoretical scenario with our simulation,
we construct estimators of the function $\psi_L$ by performing three parameter
fits of our numerical data with the analytical form (\ref{psia}).
The results for the fitted parameters $m^c$ are displayed
in Figure 9 as a function of the inverse linear lattice size $1/L$.
We observe a rather rapid variation of the estimate of the cusp
magnetization value, indicating that even for our largest lattices
the estimated function $\psi_L$ is relatively far away from its asymptotic
value. Nevertheless, the data do allow a consistent extrapolation to
the theoretical value as obtained from (\ref{circle}) (dashed curve in
the plot).

In Figure 10 we display the extrapolation of our finite volume $\psi_L$
data to their infinite volume limit. Similarly as in section 2 we assumed
the leading finite size correction to be
proportional to $1/L$. Excluding lattices of size smaller than $60$
from the fit, we obtain our infinite volume data. It is obvious from our
discussion in section 2 that a successful extrapolation of this type
determines the multimagnetical parameters. We have fitted these infinite
volume data with the analytical form of equation (\ref{psia}),
where we left the parameters $m^c$, $w$ free and fixed $m^0$ to its
exact value. We obtain a good fit consistent with the
predicted theoretical shape. The fitted value of the cusp magnetization
comes out to be $m^c=0.32(1)$, which is to be compared with the
theoretically predicted value $m^c=0.33$.

   In Figure 5 we displayed the $3D$ probability densities $p_L (m)$
at $\beta=0.2439$. We also find indications of a flat region. However
our lattices are too small to allow an estimate of the $3D$ cusp
magnetization by the above method. Therefore, we tried another approach,
which is on less solid theoretical grounds, but seems to work also for
the $3D$ Ising model, provided $\beta$ is sufficiently large. In this
approach we try to determine the cusp magnetization by measuring
the total interface area as function of the magnetization. This area
is just the number of nearest--neighbours on the lattice with opposite
spin. We assume that this area is either dominated by droplet(s) or,
after percolation, by the rectangular domain. For $D=3$, $\beta = 0.2439$
and various $L$ we plot in Figure 11 the surface area, divided by the
number of links. For large enough $L$ one can easily distinguish three
different parts in these functions. In the vicinity of of the negative
mean magnetization the surface area increases rapidly: more and more
small droplets are created. Then the droplets join to larger bubbles
and these bubbles grow in size. The increase of the total surface
area becomes then somewhat irregular, as it may decrease by joining
droplets, but over-all the increasing trend continues.
After reaching a final maximum value, the total interface area decreases
somewhat, becomes then fairly flat (very gently decreasing) until $m=0$
is reached. For positive magnetization $m$ the picture is just reflected.
The final maximum is obviously associated with percolation of the largest
bubble and the subsequent decrease has to be an entropy effect. We argue
that there is a range in the magnetization where the rectangular domain
is energetically favourable, but the free energy is still minimized by
configurations containing also some droplets. With further decreasing
magnetization the energy gap between the droplets and the domain grows
and finally the domain dominates.

To estimate the cusp magnetization $m^c$, we take the value of the
surface area at $m=0$ (where one rectangular domain dominates most), and
determine the point $m^c_L$ of the increasing part of the function with
identical surface are. This is the point where the droplets and bubbles
have the same total surface area as the domain. For those systems
where this analysis is possible the $m^c_L$ values are given in
Table 7. We cannot determine $m^c_L$ for all our systems, since
the time data series is needed, which we did not keep on disk for each
case. We notice that the cusp magnetization $m^c_L$ as function of the
lattice size is almost constant. Allowing a $1/L^{(D-1)}$ correction, we
extrapolate to $L\to\infty$. We find $m^c = 0.35 \pm 0.02$ for the $2D$
Ising model at $\beta = 0.5$. In $3D$ we find $m^c = 0.33 \pm 0.01$ at
$\beta = 0.2439$. This is in good agreement with the value one obtains
using (\ref{circle}) and our estimate $m^0 = 0.709$ of the mean
magnetization. As all our $\beta$ values are well below the roughening
point, one expect a circular shape and hence (\ref{circle}) to be valid.
\hfill\break

\section{Summary}

Multicanonical simulations allow to study the magnetic probability density
in the broken region with a hitherto unreached precision. Our numerical
calculations for the $2D$ Ising model agree well with exact results.
For the $3D$ Ising model we obtain new surface tension estimates. At
$\beta =0.232$ good agreement is found with \cite{Mey} and our amplitude
estimate agrees with Mon \cite{Mon}. When presenting the results of this
paper at the J\"ulich workshop ``Dynamics of First Order Phase
Transitions" \cite{Ha1}, we became
aware of two other recent investigations \cite{Mu1,Pot} which employ
different methods to address similar problems as the present paper.
Our results are in good agreement with \cite{Mu1} and seem also to
be consistent with \cite{Pot}.
\hfill\break

{\bf Acknowledgments:} This paper was completed at the Freie Universit\"at,
Berlin. One of the authors (BB) likes to thank H. Kleinert for his
hospitality. We are indebted to A. Billoire, K. Binder, C. Borgs, and
W. Janke for useful comments.

Our simulations were performed on the SCRI cluster of fast RISC workstations.
This research project was partially supported by
the Department of Energy under contract DE-FG05-87ER40319 and
by the Supercomputer Computations
Research Institute which is partially funded by the U.S. Department of
Energy through Contract No. DE-FC05-85ER250000.  U. Hansmann is supported
by Deutsche Forschungsgemeinschaft under contract H180411-1.
\hfill\break

\vfill\eject

\vfill\eject

\section*{Tables}

\begin{table}[pht]
\begin{center}
\caption[tab1]{The tensions $F^s_L$ as function of $L$ in $D=2$
 for $\beta = \beta_c$ ($t=0$).}
\vspace{2ex}
\begin{tabular}{||c|c|c|c|c|c||}                                        \hline
 $L$ &   No of sweeps    & $F^s_L$       & $\hat F^s_L$
                         & $\tau_L$      & $\chi_L$               \\ \hline
  2  &    $4 \cdot 10^6$ &  0.53490 (42) &          & 16.602 (1)
& 3.4974 (9) \\ \hline
  4  &    $4 \cdot 10^6$ &  0.41288 (67) &          & 90.58 (25)
& 12.195 (40)     \\ \hline
  6  &    $4 \cdot 10^6$ &  0.26958 (49) &          & 199.62 (82)
& 24.970 (13)   \\ \hline
  8  &    $4 \cdot 10^6$ &  0.20015 (66) &          & 352.6 (1.9)
& 41.379 (31)     \\ \hline
 10  &    $4 \cdot 10^6$ &  0.15891 (64) &          & 606.6 (4.4)
& 61.220 (70)      \\ \hline
 12  &    $4 \cdot 10^6$ &  0.13291 (59) &          & 828.4 (6.9)
& 84.45 (10)     \\ \hline
 14  &    $4 \cdot 10^6$ &  0.11277 (61) &          & 1083 (10)
& 110.59 (10)     \\ \hline
 16  &    $4 \cdot 10^6$ &  0.09733 (68) &          & 1446 (16)
& 139.40 (28)     \\ \hline
 18  &    $4 \cdot 10^6$ &  0.08695 (64) &          & 1846 (24)
& 171.92 (27)     \\ \hline
 20  &    $4 \cdot 10^6$ &  0.07852 (58) &          & 2424 (32)
& 206.50 (38)     \\ \hline
 30  &    $4 \cdot 10^6$ &  0.05269 (63) &          & 5253 (108)
& 422.0 (1.2)     \\ \hline
 40  &    $4 \cdot 10^6$ &  0.03922 (66) & 1.33 (32)& 10325 (281)
& 688.6 (2.4)     \\ \hline
 50  &    $4 \cdot 10^6$ &  0.03198 (68) &          & 16726 (630)
& 1021.2 (3.8)     \\ \hline
 60  &    $4 \cdot 10^6$ &  0.02620 (67) &          & 25525 (1210)
& 1412.4 (8.4)     \\ \hline
 74  &    $4 \cdot 10^6$ &  0.02098 (66) &          & 39769 (2669)
& 2016.0 (13.8)    \\ \hline
 80  &    $4 \cdot 10^6$ &  0.02050 (73) &          & 45501 (3157)
& 2355.8 (18.4)     \\ \hline
100  &    $4 \cdot 10^6$ &  0.01525 (68) & 1.33 (36)& 72877 (6720)
& 3389.5 (53.1)     \\ \hline
$\infty$& $F^s:$         &  0.00033 (34) &          &      &      \\ \hline
  exact & $F^s:$         &  0            &          &      &      \\ \hline
$\infty$& $F^s_0:$       &  $-$          & 1.33 (26)&      &      \\ \hline
  exact & $F^s_0:$       &  $-$          & 1.764... &      &      \\ \hline
\end{tabular}
\end{center}
\end{table}
\vfill\hfill\eject

\begin{table}[pht]
\begin{center}
\caption[tab1]{The tensions $F^s_L$ and effective amplitudes ${\hat F}^s_L$
in $D=2$ for $\beta = 0.47$ ($t=0.06237$).}
\vspace{2ex}
\begin{tabular}{||c|c|c|c|c|c|c||}                                    \hline
 $L$ &   No of sweeps    & $F^s_L$      & ${\hat F}^s_L$
                         & $\tau_L$      & $\chi_L$                \\ \hline
 10  &    $4 \cdot 10^6$ & 0.25705 (50) &           & 629 (5)
& 75.123 (47)  \\ \hline
 20  &    $4 \cdot 10^6$ & 0.17959 (68) & 1.94 (67) & 3516 (59)
& 292.89 (23)      \\ \hline
 40  &    $4 \cdot 10^6$ & 0.14613 (64) & 1.83 (29) & 24267 (1107)
& 1164.6 (4)      \\ \hline
 60  &    $4 \cdot 10^6$ & 0.13480 (76) & 1.76 (54) & 81338 (6701)
& 2614.1 (1.4)      \\ \hline
 80  &    $8 \cdot 10^6$ & 0.12969 (60) & 1.96 (17) & 187780 (20224)
& 4651.5 (1.4)      \\ \hline
100  &    $8 \cdot 10^6$ & 0.12814 (89) & 1.89 (36) & 571278 (80200)
& 7264.9 (2.1)      \\ \hline
$\infty$ & $F^s,m^0:$    & 0.11526 (80) &           & & 0.8522 (1) \\ \hline
exact    & $F^s,m^0:$    & 0.11492      &           & & 0.85196    \\ \hline
$\infty$ & $F^s_0:$      & 1.848 (12)   & 1.92 (17) & &            \\ \hline
exact    & $F^s_0:$      & 1.843        & 1.93      & &            \\ \hline
\end{tabular}
\end{center}
\end{table}
\vfill\hfill

\begin{table}[pht]
\begin{center}
\caption[tab1]{The tensions $F^s_L$ and effective amplitudes ${\hat F}^s_L$
in $D=2$ for $\beta = 0.5$ ($t=0.11863$).}
\vspace{2ex}
\begin{tabular}{||c|c|c|c|c|c|c||}                                    \hline
 $L$ &   No of sweeps    & $F^s_L$     & $\hat F^s_L$
                         & $\tau_L$      & $\chi_L$                \\ \hline
  2  &    $4 \cdot 10^6$ & 0.6516 (79) &             & 16.57 (5)
& 3.6692 (96) \\ \hline
  4  &    $4 \cdot 10^6$ & 0.5852 (54) &             & 94.81 (36)
     & 13.907 (61)     \\ \hline
  6  &    $4 \cdot 10^6$ & 0.4780 (06) &             & 256 (12)
     & 30.597 (10)     \\ \hline
  8  &    $4 \cdot 10^6$ & 0.4028 (04) &             & 498 (3)
    &  53.864 (148)    \\ \hline
 10  &    $4 \cdot 10^6$ & 0.3641 (04) &             & 856 (43)
    &  83.792 (32)    \\ \hline
 12  &    $4 \cdot 10^6$ & 0.3369 (14) &             & 1365 (14)
    &  120.329 (98)    \\ \hline
 14  &    $4 \cdot 10^6$ & 0.3201 (07) &             & 1909 (24)
     &  163.469 (49)    \\ \hline
 16  &    $4 \cdot 10^6$ & 0.3075 (06) &             & 2851 (45)
     &  213.330 (72)    \\ \hline
 18  &    $4 \cdot 10^6$ & 0.2991 (04) &             & 3669 (68)
     &  269.830 (62)    \\ \hline
 20  &    $4 \cdot 10^6$ & 0.2914 (07) &             & 5369 (118)
     &  332.992 (84)    \\ \hline
 30  &    $4 \cdot 10^6$ & 0.2689 (07) &             & 16688 (550)
     &  748.44 (18)    \\ \hline
 40  &    $4 \cdot 10^6$ & 0.2580 (05) &             & 37875 (1861)
     &  1329.42 (27)    \\ \hline
 50  &    $4 \cdot 10^6$ & 0.2509 (09) &             & 80526 (7562)
     &  2076.85 (60)    \\ \hline
 60  &    $8 \cdot 10^6$ & 0.2478 (07) &             & 135908 (10444)
    &   2990.66 (34)   \\ \hline
 74  &    $8 \cdot 10^6$ & 0.2443 (09) &             & 376843 (32804)
     &  4548.04 (66)    \\ \hline
 80  &    $8 \cdot 10^6$ & 0.2442 (07) &             & 585925 (83415)
   &    5315.71 (70)  \\ \hline
100  &    $8 \cdot 10^6$ & 0.2403 (12) & 2.04 (7)    & 871335 (216745)
     &  8305.01 (90)    \\ \hline
$\infty$& $F^s,m^0:$     & 0.2281(05)  &             & & 0.9113 (3) \\ \hline
  exact & $F^s,m^0:$     & 0.22806...  &             & & 0.9113     \\ \hline
$\infty$& $F^s_0:$       & 1.923 (4)   & 2.04 (7)    & &            \\ \hline
  exact & $F^s_0:$       & 1.923       & 2.10        & &            \\ \hline
\end{tabular}
\end{center}
\end{table}
\vfill\hfill\eject

\begin{table}[pht]
\begin{center}
\caption[tab1]{The tensions $F^s_L$ and effective amplitudes ${\hat F}^s_L$
in $D=3$ as function of $L$ in $D=3$  for $\beta = 0.227$ ($t=0.0236$).}
\vspace{2ex}
\begin{tabular}{||c|c|c|c|c|c||}                                      \hline
 $L$ &   No of sweeps    & $F^s_L$       & ${\hat F}^s_L$
                         & $\tau_L$      & $\chi_L$                \\ \hline
 14  &    $4 \cdot 10^6$ & 0.01146 (10) &            & 2963 (50)
& 639.41 (69)  \\ \hline
 16  &    $4 \cdot 10^6$ & 0.01099 (07) &            & 4238 (83)
& 951.1 (1.3)     \\ \hline
 20  &    $4 \cdot 10^6$ & 0.01092 (08) & 1.291 (20) & 8501 (370)
& 1853.8 (2.1)     \\ \hline
 24  &    $4 \cdot 10^6$ & 0.01113 (07) & 1.405 (69) & 14455 (576)
& 3188.7 (3.0)      \\ \hline
 26  &    $4 \cdot 10^6$ & 0.01130 (05) & 1.461 (57) & 19626 (907)
& 4046.8 (3.4)      \\ \hline
 28  &    $4 \cdot 10^6$ & 0.01149 (04) & 1.465 (71) & 24258 (1142)
& 5051 (4)     \\ \hline
 32  &    $8 \cdot 10^6$ & 0.01185 (06) & 1.538 (88) & 36323 (2570)
& 7530 (5)      \\ \hline
$\infty$  & $F^s,m^0:$   & 0.01293 (17) &            & & 0.4787 (3) \\ \hline
$\infty$  & $F^s_o:$     & 1.455 (19)   & 1.591 (62) & &            \\ \hline
\end{tabular}
\end{center}
\end{table}
\vfill\hfill\eject

\begin{table}[pht]
\begin{center}
\caption[tab1]{The tensions $F^s_L$ and effective amplitudes ${\hat F}^s_L$
in $D=3$ for $\beta = 0.232$ ($t=0.0446$).}
\vspace{2ex}
\begin{tabular}{||c|c|c|c|c|c||}                                    \hline
 $L$ & No of sweeps    &    $F^s_L$     & $\hat F^s_L$
                       & $\tau_L$      & $\chi_L$                 \\ \hline
  2  & $4 \cdot 10^6$  &  0.16391 (29) &            & 38.672 (71)
& 4.8933 (18) \\ \hline
  4  & $4 \cdot 10^6$  &  0.05297 (27) &            & 160.81 (58)
     & 26.861 (28)     \\ \hline
  6  & $4 \cdot 10^6$  &  0.03403 (15) &            & 422.1 (2.5)
     &  78.00 (8)    \\ \hline
  8  & $4 \cdot 10^6$  &  0.02779 (13) &            & 880.7 (7.6)
     & 176.15 (18)     \\ \hline
 10  & $4 \cdot 10^6$  &  0.02561 (11) &            & 1566 (19)
     & 340.76 (29)     \\ \hline
 12  & $4 \cdot 10^6$  &  0.02485 (12) &            & 2671 (41)
     & 587.10 (41)     \\ \hline
 14  & $4 \cdot 10^6$  &  0.02479 (10) &            & 4168 (82)
     & 929.26 (50)     \\ \hline
 16  & $4 \cdot 10^6$  &  0.02521 (11) &            & 6604 (151)
     & 1383.2 (9)     \\ \hline
 20  & $4.2 \cdot 10^6$&  0.02630 (11) &            & 13269 (546)
     & 2698.1 (1.8)   \\ \hline
 24  & $8 \cdot 10^6$  &  0.02752 (05) & 1.579 (20) & 23528 (1068)
     & 4652.4 (1.4)     \\ \hline
 26  & $4 \cdot 10^6$  &  0.02794 (07) & 1.613 (41) & 35032 (1985)
     & 5912.8 (2.7)    \\ \hline
 28  & $4 \cdot 10^6$  &  0.02847 (07) & 1.591 (39) & 49670 (4342)
     & 7383.7 (3.4)    \\ \hline
 32  & $6 \cdot 10^6$  &  0.02906 (08) & 1.581 (75) & 91496 (6902)
     & 11011.8 (5.3)    \\ \hline
$\infty$& $F^s,m^0:$   &  0.03140 (14) &            & & 0.5797 (1) \\ \hline
$\infty$& $F^s:$       &  1.581   (07) & 1.616 (82) & &            \\ \hline
\end{tabular}
\end{center}
\end{table}
\vfill\hfill\eject

\begin{table}[pht]
\begin{center}
\caption[tab6]{The tensions $F^s_L$ and effective amplitudes ${\hat F}^s_L$
in $D=3$ for $\beta = 0.2439$ ($t=0.0912$).}
\vspace{2ex}
\begin{tabular}{||c|c|c|c|c|c||}                                  \hline
 $L$ & No of sweeps   & $F^s_L$      & $\hat F^s_L$
                      & $\tau_L$      & $\chi_L$                 \\ \hline
  2  & $4 \cdot 10^6$ &  0.19535 (30) &            & 38.412 (51)
 & 5.2022 (22) \\ \hline
  4  & $4 \cdot 10^6$ &  0.08449 (21) &            & 193.8 (8)
     & 33.592 (23)     \\ \hline
  6  & $4 \cdot 10^6$ &  0.06691 (19) &            & 563 (4)
    & 110.058 (77)     \\ \hline
  8  & $4 \cdot 10^6$ &  0.06327 (16) &            & 1221 (12)
     & 259.85 (10)   \\ \hline
 10  & $4 \cdot 10^6$ &  0.06397 (13) &            & 2396 (35)
     & 505.738 (192)    \\ \hline
 12  & $4 \cdot 10^6$ &  0.06556 (14) &            & 4339 (81)
     & 871.35 (40)     \\ \hline
 14  & $4 \cdot 10^6$ &  0.06785 (12) &            & 7433 (196)
     & 1382.61 (50)     \\ \hline
 16  & $3.6\cdot 10^6$&  0.06989 (13) &            & 12970 (494)
     & 2061.3 (6)     \\ \hline
 20  & $4 \cdot 10^6$ &  0.07254 (09) & 1.694 (31) & 40374 (2296)
     & 4025.7 (1.5)     \\ \hline
 24  & $8 \cdot 10^6$ &  0.07349 (11) & 1.638 (53) & 76965 (7056)
     & 6954 (1)     \\ \hline
 28  & $4 \cdot 10^6$ &  0.07363 (14) & 1.599 (44) & 238560 (29158)
     & 11042 (3)    \\ \hline
 32  & $6.6\cdot 10^6$&  0.07367 (10) & 1.687 (33) & 508857 (94588)
     & 16480 (4)     \\ \hline
$\infty$ & $F^s,m^0:$ &  0.07403 (30) &            & & 0.70913 (7) \\ \hline
$\infty$ & $F^s_0:$   &  1.513 (06)   & 1.647 (59) & &             \\ \hline
\end{tabular}
\end{center}
\end{table}
\vfill\hfill

\begin{table}[h]
\begin{center}
\caption[tab1]{Cusp estimates $m^c_L$.}
\vspace{2ex}
\begin{tabular}{||c|c|c|c||}                        \hline
 $D=2$ & $\beta =0.5$ & $D=3$ & $\beta =0.2439$  \\ \hline
  $L$  & $m^c_L$      & $L$   & $m^c_L$          \\ \hline
  ~60  & 0.346 (11)   &  24   & 0.327 (07)       \\ \hline
  ~80  & 0.351 (21)   &  28   & 0.336 (14)       \\ \hline
  100  & 0.351 (45)   &  32   & 0.332 (14)       \\ \hline
\end{tabular}
\end{center}
\end{table}
\vfill\hfill\eject

\section*{Figure Captions}

{\bf Figure 1:} Probability distribution for the magnetization on
the $L=32$ lattice for $\beta=0.227$ in the $3D$ Ising model. This
distribution was obtained using the multimagnetic parameters as estimated
by our finite size scaling prediction.
\hfill\break

{\bf Figure 2:} $2D$ Magnetic probability density $P_L(M)$ for
$\beta =0.5$.
\hfill\break

{\bf Figure 3:} $2D$ effective tensions as a function of the lattice
size for $\beta =\beta_c$, 0.47, 0.5. The curves are our best fits
as described in the text.
\hfill\break

{\bf Figure 4:} $2D$ canonical heatbath and multimagnetical tunneling times
versus lattice size at $\beta =\beta_c$. The lines are power law fits and
the solid line correspond to the multimagnetical simulation.
\hfill\break

{\bf Figure 5:} $3D$ Magnetic probability density $P_L(M)$ for
$\beta =0.2439$.
\hfill\break

{\bf Figure 6:} $3D$ effective tensions as a function of the lattice
size for $\beta = 0.227$, 0.232, 0.2439. The curves are our
asymptotic fits as described in the text.
\hfill\break

{\bf Figure 7:} Typical $m\approx 0$ configuration at
$\beta = 0.5$ on a $100^2$ lattice.
\hfill\break

{\bf Figure 8:} Relying on $2D$ Ising model analytical
results \cite{RotW}: Equilibrium shapes of droplets for several
temperatures or $\beta$-values. $r(\theta)$ denotes the angle dependent radius
of the
droplet.
\hfill\break

{\bf Figure 9:} Numerical estimate of the cusp $m^c$
at $\beta =0.5$ in $2D$. The dashed line indicates a possible extrapolation
to the theoretically known infinite volume value (solid circle).
\hfill\break

{\bf Figure 10:} We plot $0.5 \cdot \psi (m)$
at $\beta =0.5$ in $2D$. The curve corresponds to a fit according to
eq. (30).
\hfill\break

{\bf Figure 11:} Surface area per link versus magnetization
at $\beta =0.2439$ in $3D$.
\hfill\break


\begin{thebibliography}{12}

\bibitem{Onsager} L. Onsager, Phys. Rev. {\bf 65}, 117 (1944).

\bibitem{RotW}  C. Rottmann and M. Wortis, Phys. Rev. {\bf B24}, 6274 (1981).

\bibitem{Shlo} S.B. Shlosman, Comm. Math. Phys. {\bf 125}, 81 (1989).

\bibitem{Bin1} K. Binder, Z. Phys. {\bf B43}, 119 (1981).

\bibitem{Bin2} K. Binder, Phys. Rev. {\bf A25}, 1699 (1982).

\bibitem{Mon} K.K. Mon, Phys. Rev. Lett. {\bf 60}, 2749 (1988).

\bibitem{Mey} H. Meyer-Ortmanns and T. Trappenberg, J. Stat. Phys.
              {\bf 58}, 185 (1990).

\bibitem{Fisk} S. Fisk and B. Widom, J. Chem. Phys. {\bf 50}, 3219 (1969).

\bibitem{Stauffer} D. Stauffer, M. Ferer and M. Wortis, Phys. Rev. Lett.
                   {\bf 29}, 345 (1972).

\bibitem{Mod} M.R. Moldover, Physical Review A {\bf 31}, 1022 (1984);
              H. Chaar, M.R. Moldover and J.W. Schmidt, J. Chem. Phys.
              {\bf 85}, 418 (1986).

\bibitem{Gie} H.L. Gielen, O.B. Verbeke and J. Thoen, J. Chem. Phys.
              {\bf 84}, 6154 (1984).

\bibitem{Fish1} M.E. Fisher and H. Wen, Phys. Rev. Lett. (to be published),
                and references given therein.

\bibitem{Widom} B. Widom, J. Chem. Phys. {\bf 43}, 3892 (1965).

\bibitem{Zin} J. Zinn-Justin, J. Physique {\bf 42}, 783 (1981).

\bibitem{Alv1} N.A. Alves, B.A. Berg and R. Villanova, Phys. Rev.
              {\bf B41}, 383 (1989).

\bibitem{Liu} A.J. Liu and M.E. Fisher, Physica {\bf A156}, 35 (1989).

\bibitem{Jan} K. Jansen, I. Montvay, G. M\"unster, T. Trappenberg, and
              U. Wolff, Nucl. Phys. {\bf B322}, 698 (1989).

\bibitem{our1} B. Berg and T. Neuhaus, Phys. Lett. {\bf B267}, 249 (1991);
               Phys. Rev. Lett. {\bf  68}, 9 (1989).

\bibitem{our2} B. Berg, U. Hansmann and T. Neuhaus, preprint,
               SCRI-91-125, submitted to J. Stat. Phys.

\bibitem{Alv2} N.A. Alves, B. Berg and S. Sanielevici, Nucl.
               Phys. {\bf B376}, 218 (1992).

\bibitem{Fer} A.M. Ferrenberg and R.H. Swendsen, Phys. Rev. Lett.
              {\bf  61}, 2635 (1988); {\bf 63 }, 1658(E) (1989), and
              references given in the erratum;
              M. Falcioni, E. Marinari, M.L. Paciello, G. Parisi and
              B. Taglienti, Phys. Lett. {\bf B108}, 331 (1982).

\bibitem{Yang} C.N. Yang, Phys. Rev. {\bf 85}, 808 (1952).

\bibitem{Bor} C. Borgs and R. Koteck\'y, J. Stat. Phys. {\bf 61}, 79
              (1990), and private communication by C. Borgs.

\bibitem{Mon2} K.K. Mon, D.P. Landau, and D. Stauffer,
               Phys. Rev. {\bf B42}, 545 (1990).

\bibitem{Wul}  G. Wulf, Z. Krist. Mineral. {\bf 34}, 449 (1901).

\bibitem{Ha1}  U. Hansmann, B. Berg and T. Neuhaus, proceedings of
               workshop ``Dynamics of First Order Phase Transitions'',
               J\"ulich, June 1--3, 1992,
               to be published in Int. J. Mod. Phys. C.

\bibitem{Mu1}  S. Klessinger and G. M\"unster, preprint, M\"unster
               University, Germany, MS-TPI-92-13.

\bibitem{Pot}  H. Gausterer, J. Potvin, C. Rebbi and S. Sanielevici,
               preprint, Boston University, BU-HEP-92-16.


\end{thebibliography}
\end{document}